\documentclass[twocolumn,showpacs,aps,prl,superscriptaddress,floatfix]{revtex4}

\usepackage{graphicx}
\begin{document}
%
%Listing the affiliations first sets the order - allows us to put CIAR last
%
\affiliation{Solid State Division, Oak Ridge National Laboratory, Oak Ridge, TN 37831}
\affiliation{Los Alamos National Laboratory, MST-10, MS K764 Los Alamos, NM 87545}
\affiliation{Department of Physics, Simon Fraser University, British Columbia Canada V5A 1S6}
\affiliation{Department of Physics and Astronomy, University of British Columbia, V6T 1Z1, Canada}
\affiliation{TRIUMF, Vancouver, British Columbia V6T 2A3, Canada}
\affiliation{Department of Physics and Astronomy, University of Tennessee, Knoxville, TN 37996}
\affiliation{Department of Chemistry, University College London, London WC1H OAJ, U.K.}
\affiliation{National Research Council, NPMR, Chalk River Laboratories,
Chalk River, Ontario, K0J 1J0, Canada}
%
%Now come the authors with their affiliation
%
\author{M. D. Lumsden}
\affiliation{Solid State Division, Oak Ridge National Laboratory, Oak Ridge, TN 37831}
\author{S. R. Dunsiger}
\affiliation{Los Alamos National Laboratory, MST-10, MS K764 Los Alamos, NM 87545}
\author{J. E. Sonier}
\affiliation{Department of Physics, Simon Fraser University, British Columbia Canada V5A 1S6}
\author{R. I. Miller}
\affiliation{Department of Physics and Astronomy, University of British Columbia, V6T 1Z1, Canada}
\author{R. F. Kiefl}
\affiliation{Department of Physics and Astronomy, University of British Columbia, V6T 1Z1, Canada}
\affiliation{TRIUMF, Vancouver, British Columbia V6T 2A3, Canada}
\affiliation{Canadian Institute for Advanced Research, 180 Dundas Street West,
 Toronto, Ontario,  M5G 1Z8, Canada}
\author{R. Jin}
\affiliation{Solid State Division, Oak Ridge National Laboratory, Oak Ridge, TN 37831}
\author{J. He}
\affiliation{Department of Physics and Astronomy, University of Tennessee, Knoxville, TN 37996}
\author{D. Mandrus}
\affiliation{Solid State Division, Oak Ridge National Laboratory, Oak Ridge, TN 37831}
\affiliation{Department of Physics and Astronomy, University of Tennessee, Knoxville, TN 37996}
\author{S. T. Bramwell}
\affiliation{Department of Chemistry, University College London, London WC1H OAJ, U.K.}
\author{J. S. Gardner}
\affiliation{National Research Council, NPMR, Chalk River Laboratories,
Chalk River, Ontario, K0J 1J0, Canada}
\title{Temperature Dependence of the Magnetic Penetration Depth  in the Vortex 
State of the Pyrochlore Superconductor, Cd$_2$Re$_2$O$_7$}
\date{\today }
\begin{abstract}
We report transverse field and zero field muon spin rotation
studies of the superconducting rhenium oxide pyrochlore,
Cd$_2$Re$_2$O$_7$.  Transverse
field measurements (H=0.007 T) show line broadening below $T_c$, which is
characteristic of a vortex state, demonstrating conclusively the
type-II nature of this superconductor.  The penetration depth is
seen to level off below about 400 mK $(T/T_c \sim 0.4)$, with a
rather large value of $\lambda (T=0) \sim 7500$~\AA.  The temperature
independent behavior below $\sim 400$ mK is 
consistent with a nodeless superconducting energy gap.
Zero-field measurements indicate no static magnetic
fields developing below the transition temperature. 
\end{abstract}
\maketitle

The pyrochlore transition metal oxides, of general formula
A$_2$B$_2$O$_7$, have been the topic of much interest in recent
years as they represent ideal systems for studying the effects of
geometrical frustration \cite{review}. Both the A and B
sublattices form a network of corner-sharing tetrahedra such that it may not be possible to energetically satisfy
all the magnetic interactions simultaneously.  The
resultant geometric frustration leads to the formation of exotic
ground states.  Much of the recent work has concentrated on local
moment systems where novel properties such as cooperative
paramagnetism \cite{Gardner_TbT}, partial, non-collinear
antiferromagnetic ordering \cite{raju,Champion_GdT}, spin-freezing
\cite{YMO}, and dipolar ``spin-ice" behavior~\cite{Harris,DyT}
have been observed.  There has, however, been growing interest in
the interplay between itinerant and local moments in geometrically
frustrated systems. The metallic pyrochlore Nd$_2$Mo$_2$O$_7$
exhibits a large anomalous Hall effect which has been attributed
to the Berry phase produced by spin chirality on the pyrochlore
lattice \cite{NdMoO}, while the spinel compound, LiV$_2$O$_4$
has been claimed to represent
the first known transition metal heavy-fermion system and
evidence exists that the unusual properties of this material are
related to geometrical frustration on the spinel lattice.
\cite{LiVO}

A vast body of work has been carried out on 3$d$ and 4$d$
transition metal pyrochlores.  These are generally insulators and
possess either a spin-glass-like or long-range ordered magnetic
structure. In contrast, the 5$d$ transition metal pyrochlores are
mainly metallic, resulting from the extended nature of the 5$d$ orbitals.  The
exception to this is Cd$_2$Os$_2$O$_7$~\cite{CdOsO}
where the 5$d^3$ configuration of Os$^{5+}$
results in a half-filled $t_{2g}$ band and a 
metal-insulator transition at 226 K.  
Despite the large number of transition metal
compounds which crystallize in the pyrochlore structure and the wide
range of physical phenomena observed in these materials, superconductivity
had not been observed until the recent discovery of bulk superconductivity
in the 5$d$ pyrochlore, Cd$_2$Re$_2$O$_7$ \cite{Sakai,Hanawa}.

Cd$_2$Re$_2$O$_7$ crystallizes in the pyrochlore structure with
room temperature lattice constant $a$=10.219~\AA~ and an oxygen
positional parameter x=0.3089.\cite{crystal}  Recent
investigations \cite{Sakai,Hanawa,Jin,Jin2,Hanawa2} have demonstrated the
existence of two phase transitions in this compound.  The first,
occurring at a temperature of about 200 K, is a continuous 
structural transition which is 
accompanied by drastic changes in resistivity and magnetic susceptibility
\cite{Jin2,Hanawa2}.
On further lowering the temperature,
Cd$_2$Re$_2$O$_7$ has been shown to exhibit bulk superconductivity
below a sample dependent transition temperature of about 1 K
\cite{Sakai,Hanawa,Jin}. Preliminary
measurements in the superconducting state indicate that
Cd$_2$Re$_2$O$_7$ is a type-II superconductor with $H_{c1}$ less
than 0.002 T and estimates of the upper critical field,
$H_{c2}$, ranging from 0.2 T to 1 T \cite{Sakai,Hanawa,Jin}. None
of the measurements reported to date extend below 0.3 K
($T/T_c\sim$0.3) and hence, little can be concluded about the
symmetry of the order parameter in this system.  An exponential
form of the specific heat as T approaches zero was speculated by
Hanawa \textit{et al.} \cite{Hanawa} but they point out that
measurements to lower temperatures are clearly needed. We report
the first measurements on Cd$_2$Re$_2$O$_7$ below 300 mK,
temperatures which are necessary (for $T_c\sim$ 1 K) to extract
information about the superconducting order parameter symmetry. We
have performed transverse field (TF) and zero field (ZF) muon spin
rotation ($\mu$SR) measurements on single crystal samples of
Cd$_2$Re$_2$O$_7$.  The ZF-$\mu$SR measurements reveal very small internal
magnetic fields which are characteristic of nuclear dipoles, indicating no
significant electronic magnetism either above or below T$_c$. The
TF-$\mu$SR results provide the first measurement of the internal field
distribution in the vortex state
in this material.  In particular, temperature dependent studies
from 20 mK to 4 K indicate a penetration depth which levels off as
$T\rightarrow 0$, suggestive of a fully gapped Fermi surface with a
rather large zero temperature value of the penetration depth,
$\lambda$(0)$\sim$7500~\AA.

Muon spin rotation has proven to be a very effective probe in the
study of superconductivity \cite{Jeff}.  In particular, TF-$\mu$SR
provides a measure of the length scales associated with type-II 
superconductors, the penetration depth, $\lambda$ and the vortex 
core radius $r_0$ \cite{Jeff}.
In a TF-$\mu$SR experiment, spin
polarized muons, with polarization perpendicular to the applied magnetic
field direction, are implanted in a sample at a location which is random on the 
length scale of the vortex lattice.  The muon precesses at a rate 
proportional to the local magnetic field providing a measure of the 
local field distribution, $n(B)$.  The presence of the vortex lattice
results in a spatially inhomogeneous field distribution and a resulting 
muon spin depolarization.

Early TF-$\mu$SR measurements assumed a Gaussian distribution of magnetic
fields and with this approximation, the penetration depth
can be directly obtained from the Gaussian depolarization rate, 
$\sigma \sim 1/\lambda^2$.  This approximation has been shown to be
reasonable for the case of polycrystalline samples but is inadequate for
the case of single crystals \cite{Jeff}.  In this case, a Ginzburg-Landau (GL)
model has been developed to model the magnetic field distribution for a
single crystal.  In GL theory,
the size of the vortex core is determined by the applied magnetic field,
$H$, and the GL coherence length normal to the applied field, $\xi_{GL}$,
while the penetration depth provides the length scale
of the decay of magnetic field away from the vortex core.  The field
distribution is calculated from the spatial distribution of magnetic
field \cite{Yaouanc},
\begin{equation}
B(\textbf{r})=\frac{\Phi_0}{S} (1-b^4) \sum_{\textbf{G}} 
e^{-i\textbf{G}\cdot\textbf{r}} \frac{u K_{1}(u)}{1+\lambda^2G^2},
\end{equation}
where $u^2=2\xi_{GL}^2 G^2 (1+b^4)[1-2b(1-b^2)]$, $K_1(u)$ is a modified
Bessel function, $\textbf{G}$ is a reciprocal lattice vector of the vortex
lattice, $b=H/H_{c2}$ is the reduced field, $\Phi_0$ is the flux quantum 
and $S$ is the area of the reduced unit cell for a hexagonal vortex lattice.

Single crystals of Cd$_2$Re$_2$O$_7$ were grown using vapor-transport
techniques as described elsewhere \cite{Jin,He}.  
Three samples with an approximate
surface area of 5$\times$5 mm$^2$ each were mounted using
low temperature grease such that the cubic (100) direction would be
parallel to the applied magnetic field direction.  They were mounted on
intrinsic GaAs in order to eliminate any precession signal at the background
frequency~\cite{kiefl85} from muons which miss the sample and would otherwise land in the Ag sample holder.  
The samples were covered with 0.025 mm Ag foil which was bolted to the 
sample holder to ensure temperature uniformity.  The TF and ZF-$\mu$SR
measurements were performed in an Oxford Instruments dilution refrigerator
on the M15 beamline at TRIUMF at temperatures from 20 mK up to 4 K.

\begin{figure}
\centering
\includegraphics[width=0.9\columnwidth,clip]{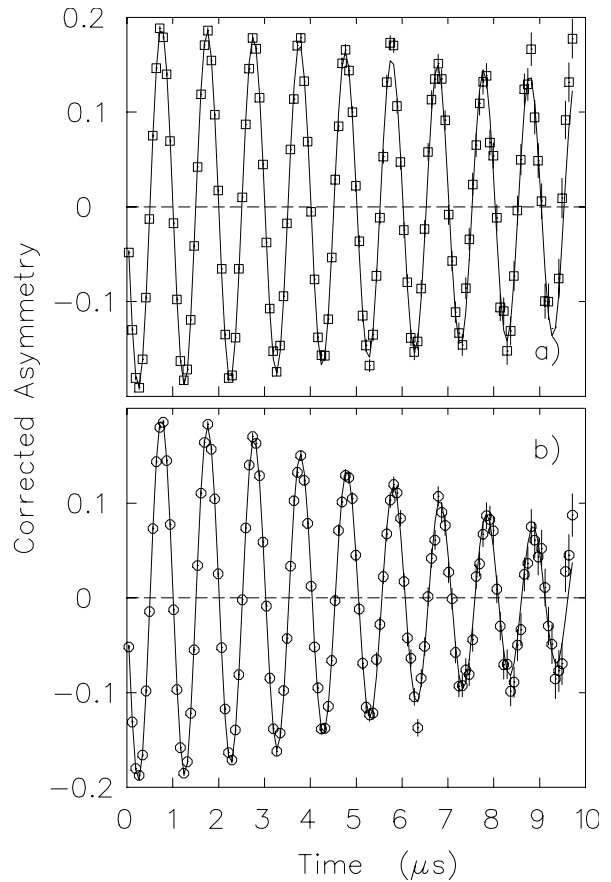}
\caption{ Typical $\mu$SR spectra in Cd$_2$Re$_2$O$_7$ obtained in
a transverse magnetic field of 0.007 T at temperatures of (a)
T=1.5 K (above T$_c$) and (b) T=100 mK (below T$_c$). }
\end{figure}

Given the estimated critical field
values, we selected a field value of 0.007 T and the temperature
dependence was measured by cooling the sample in the presence of
this applied magnetic field to ensure a uniform flux line lattice.
Figs. 1(a) and (b) show typical $\mu$SR spectra in a 
transverse field of 0.007 T, for temperatures above and below T$_c$
respectively. Examination of this data clearly shows an enhanced
depolarization rate on entering the superconducting state resulting
from the inhomogeneous field distribution associated with the flux
line lattice. This represents the first experimental observation
of the vortex lattice in Cd$_2$Re$_2$O$_7$ and provides clear
evidence that this material is a type-II superconductor.
The observed increase in the TF line broadening below T$_c$ can be 
attributed entirely to the vortex lattice since the ZF muon spin
relaxation rate (not shown here) was small and roughly temperature independent
below 2 K.

The solid lines shown in Figs. 1(a) and (b) represent fits of the
individual time spectra to a sample signal consisting of a
Gaussian envelope with fixed asymmetry and a background signal
with fixed linewidth and asymmetry.  The background, from muons which
miss the sample and land in the heat shields, was
obtained independently by performing measurements with the sample
removed.  The resulting sample linewidth, $\sigma$, is shown
in Fig. 2 as a function of temperature. As one can clearly see,
the magnitude of the depolarization rate as $T \rightarrow 0$ is very
small, saturating at a value of about 0.1 $\mu s^{-1}$. 
The residual linewidth from nuclear dipoles, taken from the data above T$_c$, 
is very small in Cd$_2$Re$_2$O$_7$ (about 0.03 $\mu s^{-1}$), 
allowing for clear observation of the line broadening associated with
the flux line lattice.  Apparently the muons stop in sites which are not close
to the Re ions, which have appreciable nuclear moments.

\begin{figure}
\centering
\includegraphics[angle=90,width=0.95\columnwidth,clip]{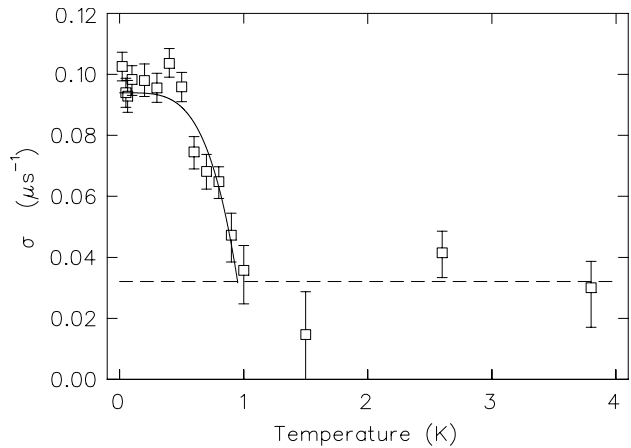}
\caption{
Linewidth parameter $\sigma$ as a function of temperature in a transverse
magnetic field of 0.007 T applied along the (100) direction.  The solid line is
a fit to Eq. 2.
}
\end{figure}

As can be seen from Eq. 1, the field distribution depends on
both the penetration depth and GL coherence depth.
The GL coherence length
 can be obtained from the known value of the upper critical
field using the expression $\xi_{GL} =(\Phi _0/2 \pi H_{c2})^{1/2}$
where $\Phi_0$ is the
flux quantum.  As mentioned above, a range of values for
$H_{c2}$ have been reported and consequently, to
provide a self-consistent measurement of $\lambda $, the field dependence
of the linewidth was measured.
To account for any possible instrumental field-dependence in the linewidth,
measurements were made above the transition temperature (2 K) at each field
value after which the sample was field-cooled to 100 mK.  
The measured linewidth in the normal state
was subtracted in quadrature from that
observed at 100 mK and the results are plotted in Fig. 3 as a function of
applied magnetic field.
As can be clearly seen, the linewidth decreases almost linearly
with applied field. This is attributed to the linear increase in the volume
taken up by the vortices.  The linewidth parameter, $\sigma_{FC}$, approaches
zero at a field of 0.5 T which is our estimate of $H_{c2}$(T$\rightarrow$0) 
and is consistent with measurements on other samples.
\begin{figure}
\centering
\includegraphics[angle=90,width=0.90\columnwidth,clip]{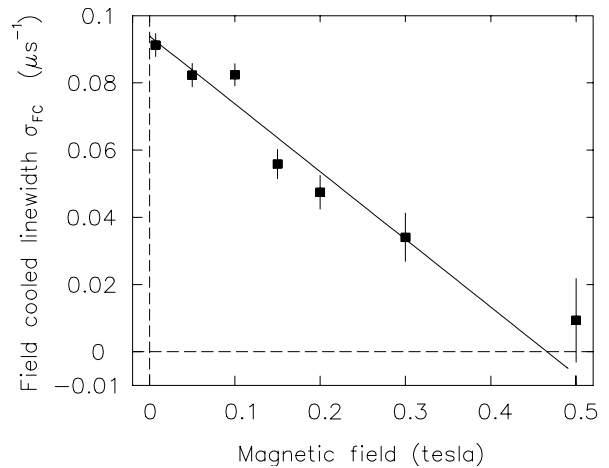}
\caption{
Linewidth parameter $\sigma_{FC}$ as a function of magnetic field applied
along the (100) direction.  The normal state contribution has been subtracted
as described in the text.  Measurements were taken at a temperature of 100 mK.
}
\end{figure}
 This estimate of the critical field corresponds to 
 $\xi_{GL} \sim$ 260~\AA. 
Using this value, we obtain the penetration 
depth using 
the field distribution shown in Eq. 1.  The
resulting temperature dependent penetration depth is shown
in Fig. 4.  As expected from the small values of linewidth, at the base
temperature we
observe a rather large value of the penetration depth,
$\lambda(0) \sim 7500$\AA.  We note that this value of
penetration depth is significantly larger than most oxide superconductors
where values ranging from 1000-2000~\AA~are typical~\cite{Uemura,Aegerter}.

\begin{figure}
\centering
\includegraphics[angle=90,width=0.9\columnwidth,clip]{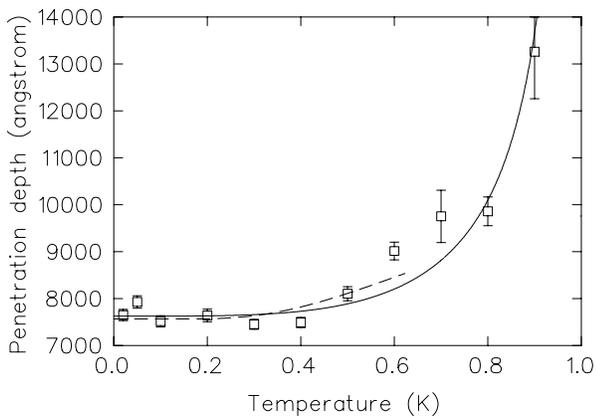}
\vspace{2ex plus 1ex minus 0.5ex}
\caption{
Penetration depth as a function of temperature in a magnetic field
of 0.007 T applied parallel to the (100) direction.  The solid line is a fit 
to Eq. 2.
}
\end{figure}

As the penetration depth is related to the concentration of superconducting
carriers, its temperature dependence is a measure of the low-lying
electronic excitations.  As such, the presence of a nodeless superconducting
energy gap is indicated by a leveling off of the penetration depth as the
temperature decreases below $T_c$.  As is clearly seen in Figs. 2 and 4, 
the linewidth and penetration depth respectively become temperature independent
 as the temperature decreases below about 0.4 K, consistent with
a fully gapped Fermi surface.  Consequently, we conclude that the
superconducting order parameter in Cd$_2$Re$_2$O$_7$ is consistent
with a nodeless energy gap suggesting either $s$-wave symmetry or
exotic pairing symmetries, such as $p$-wave which can also exhibit a
fully gapped Fermi surface.  For comparison,
the solid line in Figs. 2 and 4 represent fits to the two fluid approximation 
\begin{equation}
\frac{\sigma (T)}{\sigma (0)} \sim \frac{\lambda ^2(0)}{\lambda ^2(T)} = 
[1-(T/T_c)^4], 
\end{equation}
while the dashed line in Fig. 4 is a fit to the BCS temperature dependence
\begin{equation}
\lambda (T)= \lambda (0) \left[ 1+\sqrt \frac{\pi \Delta _0}{2 T} 
 \exp \left( \frac{-\Delta _0} {T} \right) \right]
\end{equation} 
where $\Delta_0 = 1.74(11)$ K.

The London penetration depth, $\lambda$, provides a direct measure
of the ratio of superconducting carrier concentration to effective
mass, $n_s/m^*$,
\begin{equation}
{1 \over {\lambda^2}}={{4\pi n_s e^2} \over {m^{*}c^2}}
\left( 1+{{\xi_0} \over l} \right) ^{-1},
\end{equation}
where $\xi_0$
is the Pippard coherence length, and $l$ is the mean-free path.  
There is considerable uncertainty in estimations of the mean-free path
with reported values ranging from 200-700~\AA \cite{Jin,Hiroi} and it is 
unclear whether Cd$_2$Re$_2$O$_7$ is a superconductor in the clean
or dirty limit.  
If the material is in the clean limit, the present results provide strong
evidence for a fully gapped Fermi surface.
Under the assumption of a clean superconductor, such that 
$\xi_0/l$ $\ll$ 1, a value of $n_s m_e/m^*$ of 5.0$\times$10$^{25}$ $m^{-3}$
can be obtained using Eq. 4 and the measured penetration depth.  On
the other hand, if $l\sim$200~\AA~(i.e. the dirty limit) then we obtain
$\xi_0\sim$470~\AA~and $n_s m_e/m^*$$\sim$1.4$\times$10$^{26}$ $m^{-3}$.
Clearly, precise determination of the mean free path for Cd$_2$Re$_2$O$_7$
is needed to allow accurate quantitative information to be extracted.

In conclusion, we have performed $\mu$SR studies of the superconducting
state in the recently discovered pyrochlore superconductor, Cd$_2$Re$_2$O$_7$.
Zero-field measurements indicate no significant magnetism in this 
superconductor, suggesting that magnetic frustration does not play a direct role
in the superconductivity. 
Transverse-field
measurements show that Cd$_2$Re$_2$O$_7$ is a type-II superconductor
and indicate a superconducting order parameter consistent with a fully 
gapped Fermi surface with a zero temperature value of penetration depth
of $\sim$7500~\AA.  However, considering that the superconductor may be in the
dirty limit, spectroscopic techniques which directly measure the density of
states would be required to confirm this conclusion.

We would like to acknowledge valuable discussions with M. Yethiraj, as well as
the technical support of the TRIUMF facility, in particular B. Hitti and M.
Good.
Oak Ridge National Laboratory is managed by UT-Battelle, LLC for the
U.S. Department of Energy under contract DE-AC05-00OR22725.  Work at
Los Alamos National Laboratory was performed under the auspices of the U.S.
Department of Energy.


\begin{thebibliography}{}
\bibitem{review} for recent reviews see \textbf{Magnetic Systems with Competing
Interactions}, edited by H. T. Diep (World Scientific, Singapore, 1994); A.
P. Ramirez, Annu. Rev. Mater. Sci. \textbf{24}, 453 (1994); P. Schiffer and
A. P. Ramirez, Comm. Cond. Mat. Phys. \textbf{18}, 21, (1996).
%
\bibitem{Gardner_TbT} J. S. Gardner \textit{et al.} , Phys. Rev. Lett.
\textbf{82}, 1012 (1999).
%
\bibitem{raju} N. P. Raju \textit{et al.}, Phys. Rev. B \textbf{59}, 14489
(1999).
%
\bibitem{Champion_GdT} J.D.M. Champion \textit{et al.}, Phys. Rev. B
\textbf{64}, 140407(R) (2001).
%
\bibitem{YMO}
See, for instance
N. P. Raju, E. Gmelin, R. K. Kremer, Phys. Rev. B \textbf {46},
5405 (1992); 
M.J.P. Gingras \textit{et al.}, Phys. Rev. Lett. \textbf{78}, 947, (1997).
%
\bibitem{Harris}
M.J. Harris \textit{et al.}, Phys. Rev. Lett. \textbf{79}, 2554
(1997); M.J. Harris \textit{et al.}, Phys. Rev. Lett. \textbf{81}, 4496 (1998) and
S.T.  Bramwell and M.J. Harris, J. Phys.:Condens. Matter \textbf{10}, L215 (1998).
%
\bibitem{DyT} A.P. Ramirez \textit{et al.},
Nature \textbf{399}, 333 (1999) and R.  Siddharthan \textit{et al.},
Phys. Rev. Lett. \textbf{83}, 1854 (1999).
%
\bibitem{NdMoO}
Y. Taguchi \textit{et al.}, Science \textbf{291}, 2573 (2001).
%
\bibitem{LiVO}
S. Kondo \textit{et al.}, Phys. Rev. Lett. \textbf{78}, 3729 (1997).
C. Urano \textit{et al.}, Phys. Rev. Lett. \textbf{85}, 1052 (2000).
%
\bibitem{CdOsO} D. Mandrus \textit{et al.}, Phys. Rev. B \textbf{63}, 195104 (2001).
%
%\bibitem{CaOsO} A. W. Sleight \textit{et al.},
%Solid State Commun. \textbf{14}, 357 (1974).
%
\bibitem{Sakai} 
Hironri Sakai \textit{et al.}, J. Phys. Condens. Matter \textbf{13},
L785 (2001).
%
\bibitem{Hanawa} 
M. Hanawa \textit{et al.}, Phys. Rev. Lett. \textbf{87}, 187001 (2001).
%
\bibitem{crystal}
P. Donohue \textit{et al.}, Inorg. Chem. \textbf{4}, 1152 (1965).
%
\bibitem{Jin} 
R. Jin \textit{et al.}, Phys. Rev. B \textbf{64}, 180503(R) (2001).
%
\bibitem{Jin2}
R. Jin \textit{et al.}, cond-mat/0108402 (2001).
%
\bibitem{Hanawa2}
M. Hamawa \textit{et al.}, cond-mat/0109050 (2001).
%
\bibitem{Jeff}
J.E. Sonier, J.H. Brewer and R.F. Kiefl, Rev. Mod. Phys. \textbf{72}, 769 (2000)
and references therein.
%
\bibitem{Yaouanc}
A. Yaouanc \textit{et al.}, Phys. Rev. B \textbf{55}, 11107 (1997).
%
\bibitem{He}
J. He \textit{et al.}, to be submitted.
%
\bibitem{kiefl85} R. F. Kiefl \textit{et al.}, Phys. Rev. B \textbf{32}, 530
(1985).
%
\bibitem{Uemura}
Y.J. Uemura \textit{et al.}, Phys. Rev. Lett. \textbf{66}, 2665 (1991).
%
\bibitem{Aegerter}
C.M. Aegerter \textit{et al.}, J. Phys. Condens. Matt. \textbf{10}, 7445 (1998).
%
\bibitem{Hiroi}
Z. Hiroi and M. Hanawa, cond-mat/0111126 (2001).
% 
\end{thebibliography}
\end{document}